\documentclass[a4paper, pra, twocolumn, superscriptaddress,showpacs]{revtex4-1}

\usepackage{amssymb}
\usepackage{amsmath}
\usepackage{color}
\usepackage{graphics, graphicx}
\usepackage{bbold}
\usepackage{psfrag}
\usepackage{mathcomp}
\usepackage{subfigure}
\usepackage{hyperref}

\begin{document}

\author{A. Privitera}
\affiliation{Dipartimento di Fisica, Universit\`a di Roma La Sapienza, Piazzale Aldo Moro 2, 00185 Roma, Italy}
\affiliation{Institut f\"ur Theoretische Physik, Johann Wolfgang Goethe-Universit\"at, 60438 Frankfurt am Main, Germany}
\author{W. Hofstetter}
\affiliation{Institut f\"ur Theoretische Physik, Johann Wolfgang Goethe-Universit\"at, 60438 Frankfurt am Main, Germany}


\title{Polaronic slowing of fermionic impurities in lattice Bose-Fermi mixtures}

\begin{abstract}
We generalize the application of small polaron theory to ultracold gases of Ref. [\onlinecite{jaksch_njp1}] to the case of Bose-Fermi mixtures,
where both components are loaded into an optical lattice. In a suitable range of parameters, the mixture can be described within a Bogoliubov
approach in the presence of fermionic (dynamic) impurities and an effective description in terms of polarons applies.
In the dilute limit of the slow impurity regime, the hopping of fermionic particles is exponentially renormalized
due to polaron formation, regardless of the sign of the Bose-Fermi interaction.  This should lead to clear 
experimental signatures of polaronic effects, once the regime of interest is reached. The validity of our approach
is analyzed in the light of currently available experiments. We provide results for the hopping 
renormalization factor for different values of temperature, density and Bose-Fermi interaction for three-dimensional 
$^{87}\rm{Rb}-^{40}\rm{K}$ mixtures in optical lattice.  
\end{abstract}
\maketitle
\section{Introduction}
Polaron physics and, more in general, electron-phonon interactions are one of the most influential areas  
of modern condensed matter physics

and are believed to play a major role in the physics of High-$T_c$ superconductors \cite{lanzara,alexandrov}
and strongly correlated materials. 

Ultracold gases, on the other hand, 
allow for the investigation of open issues 
in condensed matter using clean and highly tunable systems
 (see e.g. [\onlinecite{mott_esslinger}],[\onlinecite{mott_bloch}] and [\onlinecite{ferro_ketterle}]).
In the context of polaron physics, the so-called spin-polaron, i.e.  
a single spin down impurity immersed in a Fermi sea of spin up particles has been
realized as the extreme limit of imbalanced Fermi mixtures whenever $N_{\uparrow}/N_{\downarrow} \gg 1$ and 
a remarkable agreement between theory \cite{chevy,prokofev} and experiments \cite{zwierlein,nascimbene} has been achieved. 
The original polaron problem \cite{mahan} deals however with fermionic particles (electrons) interacting
with lattice vibrations (phonons), which obey the bosonic statistics, and is therefore
somehow closer to the physics of Bose-Fermi mixtures.

Bose-Fermi mixtures have been widely investigated during the last few years both 
theoretically \cite{albus,irakli1,irakli2} and experimentally \cite{eth_bosefermi,sengstock,best}.
The main focus however has been on the effect of the fermionic component of the mixture
on the coherence properties of the bosonic condensate and on the superfluid-to-Mott insulator transition.  
In addition theoretical efforts were devoted to investigate
the emergence of supersolid and other exotic phases \cite{irakli1,irakli2}. 
 
In a strongly imbalanced mixture of $N_B$ bosons and $N_F$ spinless fermions
with $N_F/N_B \ll 1$, the dilute fermionic particles act as dynamic impurities
in the bosonic condensate. On the other hand, if one focuses on the fermionic
component, the experimental setup closely resembles the polaronic problem in condensed matter, 
since fermionic atoms interact with phononic excitations of the condensate. 
Again, the main advantage of ultracold gases is that both the relative densities
of the components and their mutual interactions can be tuned much more easily 
and to a larger extent than the corresponding condensed matter
case. The extreme imbalanced limit allows for example to neglect the
interactions between different polarons and address 
the {\it{single polaron}} regime with relative simplicity.

A remarkable achievement in this direction has been, e.g., the recent experiment
performed by the Bloch group \cite{best}, where a lattice
Bose-Fermi mixture of  $^{87}\rm{Rb}-^{40}\rm{K}$ was studied, exploiting an interspecies Fano-Feshbach resonance
to tune the Bose-Fermi scattering length $a_{BF}$ and varying the relative densities $N_F/N_B$ of the species.
This allowed to study the effect of the interspecies interaction and of the population imbalance
between bosons and fermions on the transition from superfluid to Mott insulator in a very controlled way. 
 
A theoretical description of one-dimensional Bose-Fermi mixtures in terms of a 
Luttinger liquid of polarons has been proposed in [\onlinecite{mathey}], while  
the problem of polaron formation for a single impurity immersed in a homogeneous condensate 
has been studied in [\onlinecite{ciucchetti_prl06}] and more recently in [\onlinecite{tempere_prb2009}].  
Other works addressed the emergence of polarons 
in the context of cold atoms in optical lattices \cite{jaksch_prl,jaksch_njp1}.
They considered bosonic impurities loaded in an optical lattice
and the whole system immersed in a large condensate of a different bosonic species. 
Only the impurities were affected by the lattice, allowing for
an arbitrary slowing down of the impurities without perturbing the condensate
(see Section \ref{comparison} for further details). This scenario could in principle be
realized in experiments using a species-selective optical lattice. However this kind of setup,
to our knowledge, has not yet been applied to Bose-Fermi mixtures,
although several experimental schemes have been proposed \cite{leblanc} and 
species-selective lattices have already been successfully applied to Bose-Bose mixtures \cite{florence1,florence2}.  
In current experiments on Bose-Fermi mixtures in optical lattices, like e.g. in Ref. [\onlinecite{best}],
\emph{both} species are affected (though to a different extent) by the same optical lattice. In the latter case
the tunneling properties of both species are intrinsically connected to each other, and the properties of the Bogoliubov
modes of the condensate and their coupling to the fermionic particles are modified by the lattice.

For these reasons, in this work we generalize the theory developed in Ref. [\onlinecite{jaksch_njp1}] 
to the case where both the fermions and the bosons move in the same optical lattice.
Despite the presence of the lattice, we show that, in a suitable parameter regime, the bosonic
condensate still sustains phonon-like excitations and the general framework developed for the homogeneous case 
in Ref. [\onlinecite{jaksch_njp1}] still applies. We find that fermionic
particles are exponentially slowed down by the interaction with the Bogoliubov modes of the condensate, due to polaron formation. 
We also discuss the relevance of our approach to current experiments on Bose-Fermi
mixtures, analyzing the assumptions we made in order to derive our theory.  
For the specific case of a $^{87}\rm{Rb}-^{40}\rm{K}$ mixture, we provide results for the polaronic hopping renormalization
of a single fermionic impurity in several experimental setups, i.e. for different values of the bosonic density,
lattice depth, Bose-Fermi scattering length and temperature. This effect 
can actually be measured by looking at the expansion of the fermionic component of the mixture
in a lattice when the trapping potential is suddenly removed and $N_F/N_B \ll 1$ \cite{bloch_polaron}. 

The layout of the paper is the following: in the next section we explain how under
suitable conditions a Bose-Fermi mixture can be effectively described in terms of polarons 
and derive an expression for the fermionic hopping renormalization due to polaronic effects.
In Section \ref{comparison} we analyze our assumptions within a generic experimental setup
for Bose-Fermi mixtures in optical lattices. Results for the fermionic hopping renormalization 
in a $^{87}\rm{Rb}-^{40}\rm{K}$ mixture are provided in Section \ref{results}. Conclusions are drawn in Section \ref{conclusions}.  

\section{Theory}
\label{theory}
\subsection{Gross-Pitaevskii theory for static impurities}
\label{gpe_static}
The derivation of this Section follows closely the
one in Ref. [\onlinecite{jaksch_njp1}], generalizing it to
the case where  
single hyperfine states of a bosonic and fermionic species are
 loaded together into an optical lattice generated by counterpropagating
 laser beams of wavelength $\lambda$ and frequency $\omega_L$.
 For far off-resonant laser beams, the atoms experience a
 potential $V_{B/F}=V^0_{B/F}\sum_{i=1}^D \sin^2(\pi x_i/l)$ with
 $V^0_{B/F}=s_{B/F}E_r^{B/F}$, where $E_r^{B/F}=\frac{4\pi^2}{2m_{B/F}\lambda^2}\  (\hbar=1)$
is the bosonic (fermionic) recoil energy, $s_{B/F}$ denotes the dimensionless lattice depth for bosons and fermions
in the respective recoil energy and $l=\lambda/2$ is the lattice spacing, which we use as a unit length.
We choose the \emph{fermionic} recoil energy $E_r^F$ as energy unit throughout the paper.   

The effect of the trapping potential is neglected and we postpone a thorough discussion about
the correctness of our assumptions to the next section.
The system under investigation is assumed to be described 
 by a single band Bose-Fermi Hubbard model,
 with the Hamiltonian 
\begin{eqnarray}
\label{bosefermi_h}
&\hat{H}=&\hat{H}_{B}+\hat{H}_{F}+\hat{H}_{BF} \ \ \ \mbox{where}\\
&\hat{H}_B=&-J_B \sum_{<i,j>}(\hat{b}^\dagger_i \hat{b}_j+ h.c.)-\mu_B \hat{N}_B \nonumber\\
 && +\frac{U_{BB}}{2}\sum_i\hat{n}^B_i(\hat{n}^B_i-1) \\
&\hat{H}_F=&-J_F \sum_{<i,j>}( \hat{c}^\dagger_i \hat{c}_j+h.c.) -\mu_F \hat{N}_F \\
&\hat{H}_{BF}=&U_{BF} \sum_{i} \hat{b}^\dagger_i \hat{b}_i \hat{c}^\dagger_i \hat{c}_i 
\end{eqnarray}  
The fermions do not interact (directly) with each other.
 However, as we will see later, they can still interact 
via boson-mediated interactions, a situation which is very
 similar to the BCS model for standard superconductivity,
 where phonons mediate the attractive interactions between
 the electrons. We consider first the case $J_F=0$,
 where the fermions act as a set of static impurities
 on the bosonic system and their position
is specified by the (discrete) distribution function $f_i$.
 We assume that the bosonic system 
in the presence of impurities can be treated within Bogoliubov 
theory, meaning that we consider the 
solution of the Gross-Pitaevskii (GP) equation in the 
presence of impurities and quantize the oscillations around the
classical deformed ground state. If the fermionic
 impurities were absent ($f_i=0\ \ \forall i$ or equivalently $U_{BF}=0$), 
then for the unperturbed system of bosons
we can write the following GP equation by approximating
 the bosonic field operators with c-numbers
 $\hat{b}_i \approx \psi^0_i$ and $\hat{b}^\dagger_i \approx (\psi^0_i)^*$ \cite{vanoosten,burnett}, 
\begin{equation}
 \label{gp0_h}
-J_B\sum_{j \in nn_i} \psi^0_j  + U_{BB}|\psi^0_i|^2 \psi^0_i=\mu_B\psi^0_i
\end{equation}
where the sum in the first term runs over the nearest neighbors of the lattice site $i$. 
In the case of a uniform system (in the lattice), this equation is trivially solved by 
$\psi^0_i=\sqrt{n_0}$ and $\mu_B=U_{BB}n_0-zJ_B$, where $z$ is the coordination number
 of the lattice and $n_0$ is the density of particles per lattice site in the fully condensed
 state described by the classical GP theory (no quantum depletion of the condensate).
 In the presence of static fermionic impurities ($J_F=0$), the previous result has to be modified
 because the condensate macroscopic wavefunction is distorted by the impurities.
 If this distortion is sufficiently small then we can expand the classical 
field around the unperturbed solution, i.e. $\hat{b}_i \approx \psi^0_i + \delta_i$ and keep only the leading non-zero terms 
in the fluctuation $\delta_i$. This approximation is valid if $\frac{|\delta_i|}{\psi^0_i} \ll 1$ 
and in this case the GP Hamiltonian has the form 
$H_{GP}=H_0 + H_{\delta} + H_{lin}$ where :
\begin{eqnarray}
 H_0 =&& -J_B \sum_{<i,j>} [(\psi^0_i)^* \psi^0_j+h.c.] - \mu_B \sum_i |\psi^0_i|^2 \nonumber \\
&& + \frac{U_{BB}}{2}\sum_i |\psi^0_i|^4  + U_{BF}\sum_i |\psi^0_i|^2 f_i \\
H_{\delta} =&& -J_B  \sum_{<i,j>}   [\delta^*_i \delta_j +h.c.] - \mu_B \sum_i |\delta_i|^2 \nonumber\\
&&+2 U_{BB} \sum_i  |\delta_i|^2 |\psi^0_i|^2  \nonumber \\
&&+\frac{U_{BB}}{2} \sum_i (\delta^*_i |\psi^0_i|^2 \delta^*_i + \delta_i |\psi^0_i|^2  \delta_i)  \\
 H_{lin} =&& U_{BF}\sum_i [\psi^0_i \delta^*_i + (\psi^0_i)^* \delta_i]f_i       
\end{eqnarray}
where $f_i$ is the impurity distribution. 
The linear term in the fluctuation $\delta_i$ in $H_\delta$ 
is identically zero because we choose
$\psi^0_i$ as the solution of the unperturbed GP equation. 

By imposing the first derivative of this expression with respect to $\delta^*_i$ to vanish 
and using the conditions $\psi^0_i=\sqrt{n_0}$ and $-\mu_B + U_{BB}n_0=zJ$, we obtain the
 following GP equation for the fluctuation field $\delta$
\begin{equation}
 -J_B\sum_{j \in nn_i} \delta_j +zJ_B\delta_i +2U_{BB}n_0\delta_i+U_{BF}\sqrt{n_0}f_i=0
\end{equation}
This equation can be recast in the following form 
\begin{equation}
\label{gpe_latt}
\frac{\sum_{j \in nn_i} (\delta_j - \delta_i)}{l^2} - \left( \frac{2}{\xi}\right)^2 \delta_i= \frac{U_{BF}\sqrt{n_0}}{J_B l^2} f_i
\end{equation}
where we introduce the healing length of the condensate 
\begin{equation}
 \xi=\sqrt{\frac{2J_B}{U_{BB}n_0}}l
\end{equation}
Therefore the GP equation for the fluctuation field $\delta_i$ has the form of a discrete modified 
Helmholtz equation where the impurity distribution acts as a source term. For a weakly interacting condensate
the healing length $\xi$ is larger than the lattice spacing $l$ and we can consider the continuum limit 
of Eq. (\ref{gpe_latt}) applying the same considerations discussed in [\onlinecite{jaksch_njp1}]. 
The healing length $\xi$ fixes the typical scale for the variation
in space of the fluctuation field $\delta_i$ due to the impurities.
 This means that the perturbation induced in the condensate by the impurities decays exponentially in space 
with the healing length $\xi$ and the condition $\frac{|\delta_i|}{\psi^0_i} \ll 1$ (small perturbation 
of the condensate due to the impurities) implies ($U_{BB} > 0$)
\begin{equation}
\label{alpha}
 \alpha = \frac{|U_{BF}|}{U_{BB}} \frac{1}{n_0 \xi^D}=\frac{|U_{BF} |(U_{BB}n_0)^{\frac{D}{2}-1}}{(2J_B)^{\frac{D}{2}}} \ll 1 
\end{equation}
where $D$ is the dimension of the system under consideration.
Using the solution of Eq. (\ref{gpe_latt}),
the GP Hamiltonian provides the classical value of the ground state energy as a function of the impurity distribution $f_i$, i.e.
$E=E^{cl}({f_i})$.

\subsection{Bogoliubov corrections}
\label{bogo_corr}
In the previous subsection we considered how the classical condensate is distorted 
in the presence of static impurities without including any quantum effects. We now consider
the Bogoliubov excitations on top of the classical theory, decomposing the bosonic quantum operators 
in a classical and a quantum part, i.e.  $\hat{b}_i=\psi_i+\hat{\theta}_i$ where 
$\psi_i=\psi^0_i + \delta_i$ is the solution of the GP equation described above.  
If we insert this expression in the Hamiltonian and retain only the terms up to
the second order in the fluctuation fields, we obtain that all the linear terms
in the fluctuation fields (classical and quantum) disappear since we have chosen 
the classical part as the solution of the GP equation in the presence of impurities.
Therefore the Hamiltonian of the system has the form $\hat{H}=\hat{H}_{\theta} +E^{cl}({f_i})$ where 
\begin{eqnarray}
\label{bogo_hamilt}
 \hat{H}_{\theta} =&& -J_B \sum_{<i,j>}   (\hat{\theta}^\dagger_i \hat{\theta}_j +h.c.) - \mu_B \sum_i  \hat{\theta}^\dagger_i \hat{\theta}_i \nonumber\\
&&+ 2 U_{BB} \sum_i   \hat{\theta}^\dagger_i  \hat{\theta}_i |\psi^0_i|^2  \nonumber \\
&&+ \frac{U_{BB}}{2} \sum_i \left( \hat{\theta}^\dagger_i (\psi^0_i)^2  \hat{\theta}^\dagger_i +  \hat{\theta}_i (\psi^0_i)^2  \hat{\theta}_i \right)  
\end{eqnarray}
As evident from the expression above, the quantum part of the Hamiltonian is independent
of the impurity distribution and only depends on the unperturbed classical ground state
through $\psi^0_i$. The quadratic Hamiltonian $\hat{H}_\theta$ can be diagonalized using a
Bogoliubov transformation (see [\onlinecite{fetter1972}] for a general treatment) and we can re-express 
$\hat{H}_{\theta}$ in terms of the Bogoliubov modes of the condensate.

 \begin{figure}[th]
\begin{center}
\begin{tabular}{c}
\hspace{-8cm}\textbf{a)}\\
\resizebox{80mm}{!}{\includegraphics{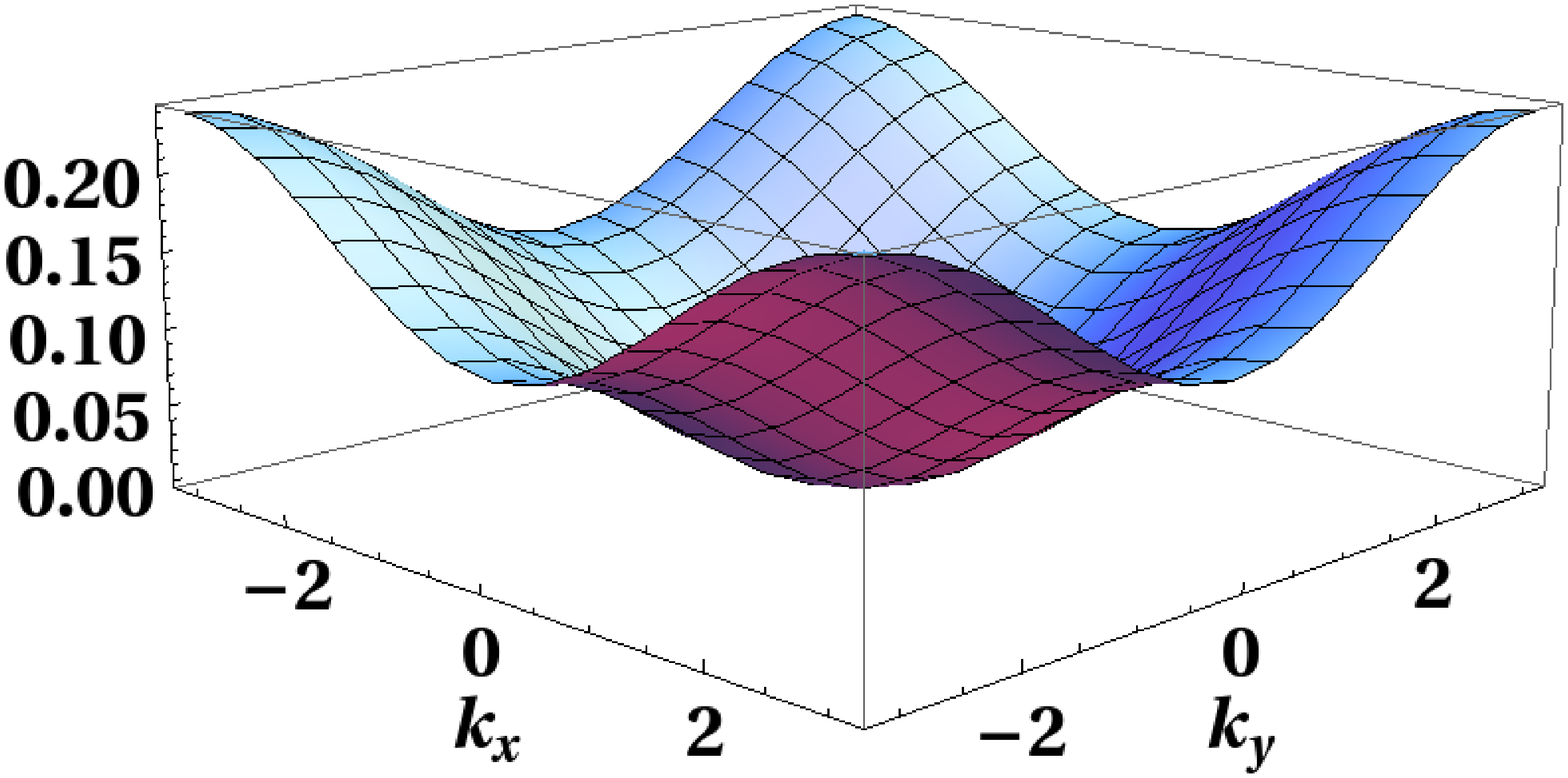}}\\
\hspace{-8cm}\textbf{b)}\\
\resizebox{80mm}{!}{\includegraphics{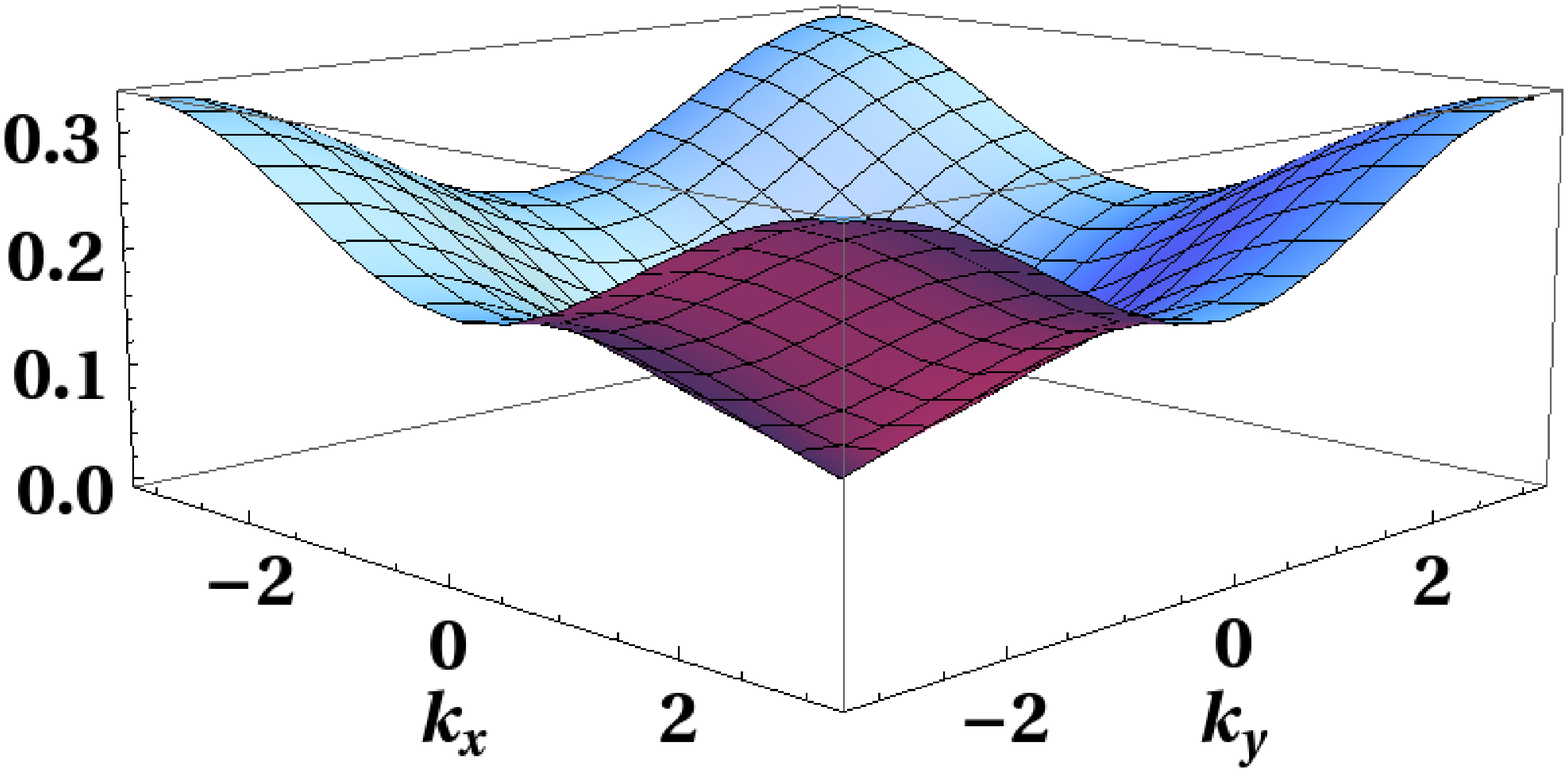}}\\
\end{tabular}
\caption{(a) Shifted single particle dispersion $\epsilon^*_{\bf{k}}$ and (b) 
Bogoliubov spectrum $\hbar \omega_{\bf{k}}$ in the $k_z=0$ plane for $D=3$ and 
 $J_B=0.029,U_{BB}=0.11,U_{BF}=0.065,n_0=1$. Energies are expressed in units of the fermionic recoil energy $E_r^F$
 and momenta in units of $l^{-1}$, where $l$ is the lattice spacing.}
\label{spectrum}
\end{center}
\end{figure}

This can be done introducing the following transformation which express the original bosonic
fluctuation operators $\hat{\theta}$ in term of new bosonic operators $\hat{\beta}$ and $\hat{\beta}^\dagger$, i.e.
\begin{equation}
\label{bogotransf}
 \hat{\theta}_i=\sum_{\mathbf{k} \in FBZ}^\prime u_{\mathbf{k},i}\hat{\beta}_{\mathbf{k}} + v_{\mathbf{k},i}^* \hat{\beta}^\dagger_{\mathbf{k}}
\end{equation}
where the sum runs over the quasimomenta $\mathbf{k}$ of the lattice within the first Brillouin zone (FBZ)
and $\mathbf{k}=0$ (the condensate) is excluded from the sum. To make $ \hat{H}_{\theta}$ quadratic the coefficients
 $u_{\mathbf{k},i}$ and $v_{\mathbf{k},i}$ have the form 
\begin{eqnarray}
&u_{\mathbf{k},i}=\frac{1}{\sqrt{N_s}}e^{i{\mathbf{k}} \cdot {\mathbf{R}}_i} u_{\mathbf{k}}
 \ \ \ \ \ \ v_{\mathbf{k},i}=\frac{1}{\sqrt{N_s}}e^{i{\mathbf{k}} \cdot {\mathbf{R}}_i} v_{\mathbf{k}}&\\
&u_{\mathbf{k}}=\sqrt{\frac{\epsilon^*_{\mathbf{k}} + U_{BB}n_0}{\hbar \omega_{\mathbf{k}}}+1}\ \ \ 
v_{\mathbf{k}}=\sqrt{\frac{\epsilon^*_{\mathbf{k}} + U_{BB}n_0}{\hbar \omega_{\mathbf{k}}}-1}&
\end{eqnarray}
where $\epsilon^*_{\mathbf{k}}=\epsilon_{\mathbf{k}}+zJ_B=2J_B\sum_{i=1}^D (1-\cos{(k_il)}) \ge 0$
 is the (shifted) single particle spectrum in tight-binding approximation, 
$\hbar \omega_{\mathbf{k}}=\sqrt{\epsilon^*_{\mathbf{k}}(\epsilon^*_{\mathbf{k}}+2U_{BB}n_0)}$
 is the energy of the Bogoliubov mode \cite{burnett,vanoosten} and $N_s$ is the number of lattice sites. 
The major difference in these expressions with respect to the continuum case treated in [\onlinecite{jaksch_njp1}]
is that the FBZ provides a natural cut-off for the single-particle energy and therefore also for the energy
of the Bogoliubov modes. In Fig. \ref{spectrum} we sketched for comparison the shifted single-particle spectrum 
$\epsilon^*_{\mathbf{k}}$ and the energy spectrum $\hbar \omega_{\mathbf{k}}$ of
 the Bogoliubov modes.

Once expressed in terms of the Bogoliubov operators, the Hamiltonian of the system is diagonal and assumes the form 
\begin{equation}
\label{hbog_stat}
 \hat{H}^{stat}= E^{cl}({f_i}) + \Delta E^{q} + \hat{H}_\beta\ \ \  \hat{H}_\beta=\sum_{\mathbf{k}
 \in FBZ}^\prime \hbar \omega_\mathbf{k} \hat{\beta}^\dagger_\mathbf{k} \hat{\beta}_\mathbf{k} 
\end{equation}
where $\Delta E^{q}$ is the quantum correction to the classical ground state energy
due to the zero-point motion of Bogoliubov modes.

 As already noticed in [\onlinecite{jaksch_njp1}],
since the Bogoliubov Hamiltonian (\ref{bogo_hamilt})  does not depend on the impurities distribution,
this means that the Bogoliubov spectrum is unaffected by the position of the impurities
and we have the same oscillation frequencies which we would have in the absence of the impurities.
The equilibrium position of these condensate oscillations are however shifted  by the presence of impurities. 
Since the Bogoliubov spectrum does not depend on the impurity positions, we can also switch the order
of the steps in the preceding derivation and calculate the Bogoliubov theory around the unperturbed ground
state (no impurities), which is given by the assumption $\hat{b}_i=\psi^0_i + \hat{\theta}_i$
and keeping only terms up to second order in the fluctuation fields $\hat{\theta_i}$. In this case we obtain
 the same expression as before for the Bogoliubov
part while the classical part is not anymore the solution of GP equation in the presence of impurities. The Hamiltonian operator now has the form 
$\hat{H}=E_{\psi^0}+\hat{H}_\theta+\hat{H}_{lin}$, where 
\begin{equation}
  H_{lin} = U_{BF}\sqrt{n_0} \sum_i  f_i [\hat{\theta}^\dagger_i + \hat{\theta}_i ] 
\end{equation}
and $E_{\psi^0}$ is a c-number. Using the Bogoliubov transformation (\ref{bogotransf}),
we get in terms of the Bogoliubov modes  
 \begin{eqnarray}
\label{holstein1}
\hat{H}&=&E_{\psi^0}+\hat{H}_\theta+\hat{H}_{lin}\\
\label{holstein2}
 H_{\theta} &&=  \sum_{\mathbf{k} \in FBZ}^\prime \hbar \omega_{\mathbf{k}} \beta^\dagger_{\mathbf{k}}\beta_{\mathbf{k}} \\
\label{holstein3}
H_{lin}&& = \sum_i  \sum_{\mathbf{k}\in FBZ}^\prime  \hbar \omega_{\mathbf{k}} [M_{i,{\mathbf{k}}}\beta_{\mathbf{k}}  + M^*_{i,{\mathbf{k}}} \beta^\dagger_{\mathbf{k}}]f_i 
\end{eqnarray}
where 
\begin{equation}
\label{m_lat}
 M_{i,{\mathbf{k}}}=\frac{U_{BF}\sqrt{n_0}}{\hbar \omega_{\mathbf{k}}}(u_{\mathbf{k},i}+v_{\mathbf{k},i})=
M_{\mathbf{k}}e^{i{\mathbf{k}} \cdot {\mathbf{R}}_i} 
\end{equation}
and 
\begin{equation}
 M_{\mathbf{k}}= U_{BF}\sqrt{\frac{n_0 \epsilon^*_{\mathbf{k}}}{N_s(\hbar \omega_{\mathbf{k}})^3}}
\end{equation}

In this case the new bosonic operators annihilate Bogoliubov excitations around 
the unperturbed ground state and therefore do not annihilate the real vacuum defined above
 which is distorted by the impurities. 
This leaves us with a generalized Holstein model with phonons coupled
to the fermionic density \cite{holstein1,holstein2}. The main difference with respect to the 
original Holstein model is that here we have a \emph{continuum}
 of phonons instead of a single phononic mode.
The high-energy phonon contribution in this expression is cut-off due to the FBZ,
while in the continuum case \cite{jaksch_njp1} the physical cut-off is
provided by the inverse of the typical localization length of the impurities 
in the Wannier states, which appears explicitly in the matrix elements $M$
(see subsection \ref{single_imp}). For static impurities this Hamiltonian can be diagonalized
 by introducing a unitary Lang-Firsov \cite{langfirsov} transformation
which shifts the equilibrium position of the condensate around the places where the impurities are localized 
\begin{equation}
\label{lang-firsov}
 \hat{U}=\exp{\left[ \sum_{j}\sum_{{\mathbf{k}} \in FBZ}^\prime \left( M^*_{j,{\mathbf{k}}} \hat{\beta}^\dagger_{\mathbf{k}} - M_{j,{\mathbf{k}}} \hat{\beta}_{\mathbf{k}}  \right) \right]}f_j  
\end{equation}
 This makes the Hamiltonian diagonal in the bosonic operators recovering Eq. (\ref{hbog_stat}). 
\subsection{Slowly moving impurities}
\label{slow_mov}
If $J_F$ is not zero, i.e. if the impurities can move through the lattice, 
then the problem is not fully solved using (\ref{lang-firsov}).
However the Lang-Firsov transformation provides physical insight on how to proceed.
Indeed, introducing now the fermionic operators $\hat{c}_i$ and $\hat{c}_i^\dagger$, we can repeat the steps above 
by simply replacing the density distribution $f_i$ with the density operator $\hat{n}_i=\hat{c}_i^\dagger \hat{c}_i$ everywhere.
The Lang-Firsov transformation now acts simultaneously on fermionic and bosonic degrees of freedom
\begin{equation}
 \hat{U}=\exp{\left[ \sum_{j}\sum_{{\mathbf{k}}\in FBZ}^\prime \left( M^*_{j,{\mathbf{k}}}
 \hat{\beta}^\dagger_{\mathbf{k}} - M_{j,{\mathbf{k}}} \hat{\beta}_{\mathbf{k}}  \right) \right]}\hat{n}_j  
\end{equation}
Using the Baker-Hausdorff formula it is possible to show that $\hat{U}\hat{\beta}^\dagger_{\mathbf{k}} 
\hat{U}^\dagger=\hat{\beta}^\dagger_{\mathbf{k}} -\sum_j M_{j,{\mathbf{k}}} \hat{n}_j$,
 $\hat{U}\hat{n}_j \hat{U}^\dagger=\hat{n}_j$ and $\hat{U}\hat{c}^\dagger_j \hat{U}^\dagger=\hat{c}^\dagger_j \hat{X}^\dagger_j$ where the 
operator $\hat{X}^\dagger_j$ creates a coherent cloud of Bogoliubov modes around the position j, i.e.
\begin{equation}
 \hat{X}^\dagger_j=\exp{\left[  \sum_{{\mathbf{k}}\in FBZ}^\prime \left( M^*_{j,{\mathbf{k}}} \hat{\beta}^\dagger_{\mathbf{k}} - M_{j,{\mathbf{k}}} \hat{\beta}_{\mathbf{k}}  \right) \right]}
\end{equation}
The Lang-Firsov transformed Hamiltonian now has the form 
\begin{eqnarray}
\label{h_hol_lftransf}
  \hat{H}_{LF} &&=
  -J_F \sum_{<i,j>} (\hat{X}_i \hat{c}_i)^\dagger (\hat{X}_j \hat{c}_j) 
- \tilde{\mu} \hat{N}_F  -\frac{1}{2}\sum_{i\neq j} V_{i,j} \hat{n}_i  \hat{n}_j \nonumber \\
&&+ \sum^\prime_{\mathbf{k}\in FBZ}  \hbar \omega_{\mathbf{k}} \hat{b}^\dagger_{\mathbf{k}} \hat{b}_{\mathbf{k}}
+ E  
\end{eqnarray}
As already pointed out, the presence of Bogoliubov modes induces (offsite) interactions between the impurities.
The interaction potential has the form 
\begin{equation}
 V_{i,j}=2\sum^\prime_{\mathbf{k}\in FBZ} \hbar \omega_{\mathbf{k}} |M_{\mathbf{k}}|^2 \cos{
\left[{\mathbf{k}} \cdot ({\mathbf{R}}_i-{\mathbf{R}}_j)\right]}
\end{equation}
 
\begin{figure}
\includegraphics[scale=0.25]{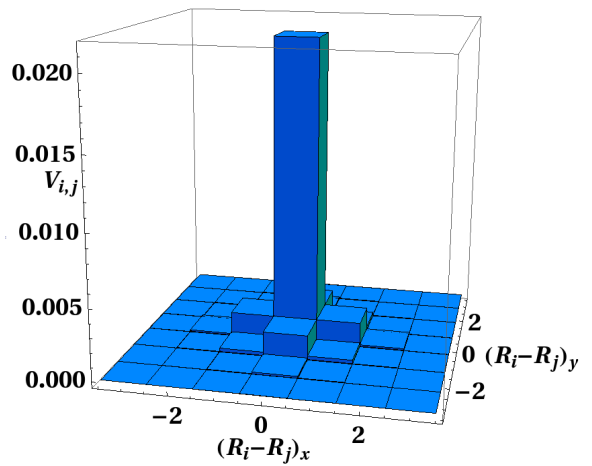}  
\caption{(Color online) Effective interaction potential $V_{i,j}$
 between the impurities for $({\mathbf{R}}_i-{\mathbf{R}}_j)_z=0$ and $D=3$ ($J_B=0.029,U_{BB}=0.11,U_{BF}=0.065,n_0=1$).
 The central peak for $i=j$ is proportional to the polaron shift ($V_{i,i}=2 E_p$).
 Energies are expressed in units of $E_r^F$ and lengths in units of $l$.}
\label{potential}
\end{figure}

Therefore even though spin-polarized fermionic impurities do not interact directly with each other, 
their coupling to the Bogoliubov modes of the condensate creates an effective interaction between them.
As shown in Fig. (\ref{potential}), the interaction is always attractive ($V_{i,j} > 0$) 
regardless of the sign of the Bose-Fermi interaction and decays very fast
with the distance between the impurities. Moreover, during its motion, the impurity drags a Bogoliubov cloud 
and this affects its kinetic energy, as evident from the first term in Eq. (\ref{h_hol_lftransf}).
If the impurities are not moving ($J_F=0$), their energies are lowered by
an amount of energy which represents the potential energy gain due to the
interaction with the Bogoliubov cloud. This characteristic energy scale
for static impurities is the polaron shift $E_p$, where
\begin{equation}
\label{polaron_shift}
 E_p=\sum_{\mathbf{k} \in FBZ}^\prime \hbar \omega_{\mathbf{k}} |M_{\mathbf{k}}|^2= U_{BF}^2 \frac{1}{N_s}\sum_{\mathbf{k} \in FBZ}^\prime
 \frac{n_0 \epsilon^*_{\mathbf{k}}}{(\hbar \omega_{\mathbf{k}})^2} 
\end{equation}
and $\tilde{\mu}=\mu_F+E_p$ in Eq. (\ref{h_hol_lftransf}). 
 
Whenever this energy scale is much larger than the hopping parameter of the impurities, i.e. 
\begin{equation}
\label{slow_imp}
\zeta=\frac{J_F}{E_p} \ll 1
\end{equation}
we expect that the impurity and its surrounding cloud will tunnel together like a composite object, i.e. they form a
\emph{polaron}. The expressions (\ref{holstein1},\ref{holstein2},\ref{holstein3}) and their Lang-Firsov transformed version (\ref{h_hol_lftransf}) 
are in practice valid even beyond the condition (\ref{slow_imp})  (slow impurity regime), 
while the results we derive below assume that impurities are slow in the sense specified by Eq. (\ref{slow_imp}).   
 In this sense this treatment is analogous to the the antiadiabatic limit (fast phonons) of the small polaron theory introduced
 by Holstein \cite{holstein1,holstein2} where the impurities move much slower than the time taken by the coherent cloud to rearrange itself. 
Here we would like to point out that in the expression (\ref{slow_imp}), the bare fermionic hopping $J_F$ 
is not compared with the bare bosonic hopping $J_B$ but with $E_p$ which explicitly depends also on the Bose-Bose interaction and the bosonic density.
This is because in Bose-Fermi mixtures the \emph{excitations} of the bosonic condensate and not the original bosonic particles play the 
role analogous to of phonons in the standard polaronic problem.

\subsection{Single fermionic impurity in the strong coupling small polaron regime}
\label{single_imp}
We consider now the case of a single fermionic impurity immersed in a much larger BEC. In real experiments this \emph{single polaron} regime 
is realized whenever $N_B \gg N_F$ such that $(n_F)^{-1/D} >> \xi/l$, where $n_F=N_F/N_s$ is the number of fermions per lattice site. 
This implies that the average interparticle distance is much larger than the healing length of the BEC, 
so that also  interactions induced by the Bogoliubov modes can be neglected.
For a single impurity the Lang-Firsov Hamiltonian becomes 
\begin{equation}
H_{1-imp}=\sum_{\mathbf{k} \in FBZ}^\prime  \hbar \omega_{\mathbf{k}} \beta^\dagger_{\mathbf{k}}\beta_{\mathbf{k}} -J_F \sum_{<i,j>} (\hat{X}_i \hat{c}_i)^\dagger (\hat{X}_j \hat{c}_j)) + E_0
\end{equation}
where $E_0$ is a c-number. For $J_F=0$ the fermionic and bosonic part are completely disconnected and the impurity can sit everywhere in the lattice
with the same energy. If $J_F$ is nonzero but $\zeta=\frac{J_F}{E_p} \ll 1$ the polaron is the appropriate quasiparticle and 
the hopping term can be treated as a small perturbation. We focus now on the regime of temperature $k_B T \ll E_p $, 
where incoherent phononic scattering is highly suppressed \cite{jaksch_njp1}. The degeneracy of the Wannier states
can be removed by introducing Bloch waves labeled by $\mathbf{k}^\prime$ for the impurity and considering 
\begin{equation}
\Delta E (\mathbf{k^\prime},\{N_{\mathbf{k}}\})\!=\!\langle \mathbf{k}^\prime,\{N_{\mathbf{k}}\}|-J_F\!\!\!
\sum_{<i,j>}\!\!(\hat{X_i}\hat{c_i})^\dagger\!(\hat{X_j}\hat{c_j})|\mathbf{k}^\prime,\{N_{\mathbf{k}}\} \rangle 
\end{equation}
where $\{N_{\mathbf{k}}\}$ indicates the configuration of Bogoliubov modes. This matrix element can be calculated
using standard techniques for phonons \cite{mahan}. If we assume thermally distributed phonons, we get 
that the bare hopping of the impurity $J_F$ is exponentially renormalized to $J_F^r=J_F e^{-S}$
and the renormalization factor $S$ is given by
\begin{equation}
 S=\sum_{\mathbf{k} \in FBZ}^\prime |M_{0,{\mathbf{k}}}|^2 [1-\cos({\mathbf{k}} \cdot {\mathbf{a}})] (2N_{\mathbf{k}}+1)
\end{equation}
 where $N_{\mathbf{k}}=\frac{1}{e^{\hbar \omega_{\mathbf{k}}/k_B T}-1}$. 
Inserting the expression  (\ref{m_lat}) of the matrix elements $M$, one obtains that $S(T,U_{BF})=U^2_{BF}f(T)$, where
\begin{equation}
\label{f_factor}
  f(T)=\frac{1}{N_s}\sum_{\mathbf{k} \in FBZ} 
\frac{n_0 \epsilon^*_{\mathbf{k}}}{(\hbar\omega_{\mathbf{k}})^3}
[1-\cos(\mathbf{k}\cdot \mathbf{a})]
\ (2N_{\mathbf{k}}(T) +1) 
\end{equation}

In the practice, within this approach, the renormalization factor $S$ is proportional to the square
of the Bose-Fermi interaction $U_{BF}$, while the factor $f$ only depends on the condensate properties.
As evident from the expression above, $S$ does not depend on the sign of the
Bose-Fermi interaction. This results in a Gaussian dependence of the renormalized hopping on
the Bose-Fermi interaction, i.e. $J_F^r=J_F e^{-U^2_{BF}f}$.
The prefactor $f$, together with its dependency on the temperature $T$ and on other parameters
like $U_{BB},J_B$ and $n_0$, can then be calculated independently and results will be
presented in Section \ref{results} for the $^{87}\rm{Rb}-^{40}\rm{K}$ case.

\begin{figure}
\includegraphics[scale=0.35]{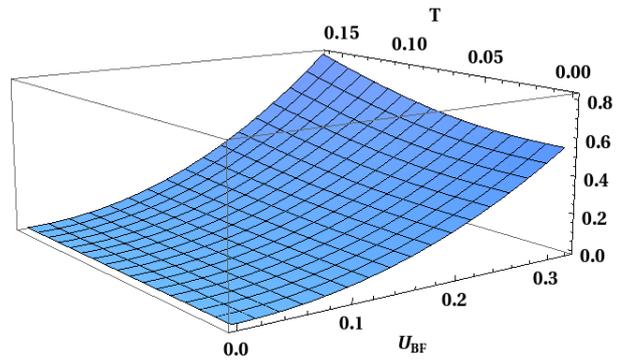}  
\vspace{-0.4cm}
\caption{(Color online) Sketch of the renormalization factor $S$ as a function of Bose-Fermi coupling $U_{BF}$
 and temperature $T$ ($k_B=1$) for fixed $n_0$. Energies are expressed in units of $E_r^F$.}
\label{damping}
\end{figure}

All the results presented in this Section are expressed in terms of the parameter $n_0$, the density of particles in the condensate.
However a fixed value of $n_0$ corresponds to different values of the bosonic density $n_B$, whenever 
one changes the temperature $T$ or the Hamiltonian parameters $U_{BB},J_B$. In order to compare
data with experiments, the results have to be expressed in terms of the bosonic density $n_B$ and 
$n_0(T)_{|n_B}$ has to be calculated selfconsistently, as explained in the next section. 
This point is crucial for the understanding of the temperature dependence of the $S$ factor.
For example, for fixed values of $n_0$, $S$ increases with temperature due to the increasing number of excited
phonons, and therefore one would expect that the minimal slowing of the impurities occurs at $T=0$, where
only the zero point motion of the Bogoliubov modes contributes. At the same time however, $n_0$ decreases
with $T$ for fixed bosonic density $n_B$ and the overall temperature dependence is determined by the competition between
thermal depletion of the condensate and thermal excitation of the Bogoliubov modes. The energy 
spectrum of the Bogoliubov modes $\hbar \omega_{\bf{k}}$ acquires a temperature dependence
through $n_0(T)_{|n_B}$ which is missing in the standard condensed matter case,  
where an increasing temperature only increases the phononic population and therefore the $S$ factor \cite{holstein2}.
This results in a mayor difference between the condensed matter and present case, and 
also suggests the existence of a richer temperature dependence of polaronic effects in the Bose-Fermi mixtures 
realization. Indeed, it is possible that different mixtures (or even the same mixture for different parameter range)
 show different slopes in $S(T)$ or even a nonmonotonic behavior. 
 A sketch of the dependence of the renormalization factor $S$ on $T$ and 
$U_{BF}$ for fixed $n_0$ is drawn in Fig. \ref{damping}. This would be the relevant case whenever the 
depletion can be neglected in the range of parameters under investigation and $n_0 \approx n_B$. 
As shown in Section \ref{results}, this is not the case for the $^{87}\rm{Rb}-^{40}\rm{K}$ setup we considered.   
In any case it is worth noting that also at $T=0$ there is a sizable contribution to the S factor 
from all the Bogoliubov modes with quasimomenta within the FBZ. 

Whenever  the condensate is unaffected by the lattice, e.g. in the case considered in [\onlinecite{jaksch_njp1}],
the Bogoliubov modes are labeled like plane waves and the sum runs over all the possible
momenta $\mathbf{q}$. In the lattice the FBZ provide a natural cut-off to the high-energy phonons contribution,
while the analogous role of the physical cutoff in the continuum case is played by an additional exponential decay
of the matrix elements $M$ for large momenta. Indeed the matrix elements $M$ in the continuum case \cite{jaksch_njp1}
are given by 
\begin{equation}
M_{j,{\mathbf{q}}} \propto \sqrt{\frac{n_0\epsilon_{\mathbf{q}}}{(\hbar \omega_{\mathbf{q}})^3}}f_j({\mathbf{q}})
\end{equation}
where $\epsilon_{\mathbf{q}}$ is the free-particle dispersion, $\hbar \omega_{\mathbf{q}}$ is the energy of the Bogoliubov
modes in the continuum \cite{jaksch_njp1},
\begin{equation}
\label{f_factor_continuum}
f_j({\mathbf{q}})=\int d{\mathbf{r}} |\chi_j({\mathbf{r}})|^2 \exp{(i{\mathbf{q}}\cdot{\mathbf{r}})} 
\end{equation}
and $\chi_j({\mathbf{r}})$ is the Wannier wavefunction of the impurity localized at site $j$. 
Therefore $M$ is proportional to the Fourier transform with respect to the momentum $-{\mathbf{q}}$
of the density profile of the impurity in a Wannier state. Using a Gaussian approximation of the Wannier
wavefunction, Eq. (\ref{f_factor_continuum}) reduces to   
\begin{equation}
 f_j({\mathbf{q}}) \approx (\exp{(-\frac{q^2\sigma^2}{4})})^d\exp{(i{\mathbf{q}}\cdot{\mathbf{r}}})
\end{equation}
where $\sigma$ is the width of the Gaussian wavefunction. Therefore the modulus of the matrix element $M$
decays exponentially for $q \ll q_c=1/\sigma$. In a deep lattice typically $\sigma \ll l$ and 
the cut-off is much larger than the Brillouin zone. Therefore the continuum theory cannot be used
to address quantitatively the lattice case, not even by introducing an effective mass $m_e$ to
take the lattice in account. On the other hand a situation in which 
the bosonic condensate is unaffected by the lattice would be more favorable to realize the \emph{slow impurity}
regime described above, since the impurity can be slowed down arbitrarily without affecting the bosonic kinetic energy. 
In current experiments on Bose-Fermi mixtures in a lattice this is not the case as explained in the next section
and the hopping parameters $J_F$ and $J_B$ are related to each other.

\section{Limits of validity of the theory}
\label{comparison}
In this Section we analyze the assumptions we made in deriving our approach
within a generic experimental setup for Bose-Fermi mixtures in optical lattices.  

Since bosons and fermions are loaded into a single optical lattice of frequency $\omega_L$, both species experience
the same laser intensity $I$ and $V^0_{B/F}=\alpha_{B/F}(\omega_L) I$,
where $\alpha(\omega_L)$ is the atomic dynamic polarizability at the laser frequency.
We define the dimensionless parameter 
\begin{equation}
 \gamma=\frac{s_B}{s_F}=\frac{\alpha_B(\omega_L)m_B}{\alpha_F(\omega_L)m_F} 
\end{equation}
which rules the ratio between the lattice depth experienced by bosons and fermions
and therefore between their kinetic energies. For fixed atomic species, $\gamma$ can be
varied by changing the wavelength of the optical lattice. 

 In the presence of a interspecies Fano-Feshbach resonance, the Bose-Fermi scattering length $a_{BF}$ can be 
tuned by a magnetic field $B$, 
while the Bose-Bose scattering length $a_B$ can be considered in practice as constant in the same range of $B$.
 As already discussed previously, polarized fermions do not interact directly with each other.

In a real experiment, the atomic gas is confined in a trapping potential, which is not included
 in our approach. In the three-dimensional case and for sufficiently shallow trapping potentials, 
our results can still be considered locally in a LDA framework. The situation is very different in $D=2$.
In this case the trap radically modifies the properties of the system, providing a cut-off to the long-wavelength
Goldstone modes, which would destroy the condensate at any finite temperature in the homogeneous case. 
Therefore, even though formally our approach to the homogeneous setup in the two-dimensional case
is well defined at $T=0$, the Bogoliubov treatment of the bosonic component immediately breaks down 
at any finite temperature in that case, and therefore we cannot directly apply our findings to a two-dimensional setup. 

In practice (quasi) two-dimensional systems are obtained by strongly increasing the optical 
lattice in one direction (e.g. in the $z$ direction) such that the motion in this direction is frozen and only the 
zero point motion has to be considered \cite{hadzibabic}. Under this condition it is possible to distinguish between two regimes.
Whenever $a_B \ll \sigma_z$, where $\sigma_z$ is the typical width of the onsite wavefunction in the direction of 
the tight confinement, then the scattering process is still essentially three-dimensional even if the motion is essentially two-dimensional.
In this case the $3D$ scattering length can be safely used and the local interaction $U$ increases with the confinement 
in the $z$ direction since the overlap between the local wavefunctions is increased. 
For the case $a_B \leq \sigma_z$ a more refined treatment is required \cite{petrov_2}. 
In this case strong modifications of the interaction both in modulus and sign can occur in the system.    
The main feature of interest for the present paper is that the confinement  
can actually be used to further tune the interaction between the bosons
in order to access different regimes in Bose-Fermi mixtures. 
Even though the existence of a real condensate has been predicted  for the case of trapped (quasi-) two-dimensional
 setups \cite{petrov_1,gies} at low temperature,
the application of the Bogoliubov approach necessarily needs the trapping potential to be explicitly included in the treatment. 
Since in this case a full numerical solution of the corresponding Bogoliubov theory is required,
we postpone the analysis of this interesting case to the future despite its intrinsic interest.
On the other hand we expect the general conclusions of the paper to be still
valid also in a $2D$ setup. For the reasons above, the approach is developed in the general case, when possible,
but only results for the three-dimensional case are shown.      
 
\subsection*{Single band Bose-Fermi Hubbard model}
Our first assumption is that the Bose-Fermi mixture under investigation is described using
 the single-band Bose-Fermi Hubbard model (\ref{bosefermi_h}).
This requires that i) higher bands contributions, ii) non-local interaction terms, 
and iii) next-nearest neighbors hopping terms are negligible in the parameter range under investigation. 
The first condition is particularly crucial for the bosonic component where the local density can take arbitrarily large values. For deep enough optical lattices, a Gaussian approximation can be used 
to estimate onsite parameters for our model. In this approximation $W_j({\mathbf{x}})=\prod_{i=1}^D W^G_j(x_i)$ where  $W^G_j(x_i)$ is a Gaussian wavefunction in one dimension ($x_i=x,y,z$ for $D=3$) 
localized around the site $j$ of the optical lattice, i.e. 
\begin{equation}
W^G_j(x_i)=(\pi\sigma^2)^{-1/4}\exp{[-(x_i-R_j)^2/(2\sigma^2)]}                                      
\end{equation}
where $\sigma_{B,F}=\sqrt{\hbar/m_{B,F}\omega_{B,F}}$ and $\hbar \omega_{B,F}=2E_r^{B,F}\sqrt{s_{B,F}}$. 
In the same approximation 
\begin{eqnarray}
 U_{BB}&=&\frac{4\pi a_B}{m_B} \left(\frac{\pi^2 s_B}{4}\right)^{D/4}\\
 U_{BF}&=&\frac{2\pi a_{BF}}{m_r} \left(\frac{\pi}{s_B^{-1/2} + s_F^{-1/2}}\right)^{D/2}
\end{eqnarray}
where we have set $(E_r^F=\hbar=l=1)$ and $m_r=1/(m_B^{-1}+m_F^{-1})$, and a $\delta$-like pseudopotential
has been used to model the interaction between particles. 
The Gaussian approximation provides a very poor estimate of the hopping parameter,
 which can be expressed in a simple way for deep enough lattices (large $s$)
using the asymptotic solution of the Mathieu equation \cite{zwerger}
\begin{equation}
J_{B/F}/E^{B/F}_r = \frac{4 s_{B/F}^{3/4}}{\sqrt{\pi}}\exp{[-2\sqrt{s_{B/F}}]}
\end{equation}
In practice however this formula applies with reasonable accuracy only for $s \geq 10$, while for smaller $s$ values the hopping term 
is overestimated and a direct numerical evaluation of the hopping parameter is required. Consistency with the model (\ref{bosefermi_h})
requires \cite{albus} 
\begin{equation}
a_{BF},a_B \ll \sigma_{B,F} \ll l \ \ \&\ \  \frac{U_{BB}}{2}n_B(n_B-1) \ll \hbar \omega_B
\end{equation}
These conditions are reasonably well satisfied in practice if the onsite density of bosons $n_B$ is not too large.  
For increasing lattice depth $s$, the hopping parameter $J$ decreases exponentially, while the interaction term is slightly increased because 
of the increasing onsite overlap of the Wannier orbitals. It is important to point out that since both species move 
in the same optical lattice, the following relation applies for large $s_F$ and fixed $\gamma$:
\begin{equation}
 \frac{J_F}{J_B}=\frac{m_B}{m_F}\gamma^{-(3/4)}\exp{[(2*\sqrt{s_F})^{(\sqrt{\gamma} - 1)}]}
\end{equation}
and therefore the ratio $J_F/J_B$ is not constant for fixed $\gamma$ but still depends on the lattice depth $s_F$.  

\subsection*{Bogoliubov approach}
Our approach is based on the possibility of describing the bosonic component of the mixture
in the presence of static or slowly moving impurities within Bogoliubov approach.
This requires in general that neither quantum nor thermal fluctuations 
are strong enough to substantially deplete the condensate, i.e. the condensate fraction 
\begin{equation}
\label{cond_frac}
 \phi=N_0/N_B \leq 1
\end{equation}
 needs to be close to 1. 

The parameter $\alpha$ introduced in Eq. (\ref{alpha}) quantifies the effect of the impurities
on the condensate wavefunction, such that if $\alpha \ll 1$ we can expand the GP equation around
the unperturbed solution in the absence of impurities. It is worth mentioning that $\alpha$ is markedly
dependent on the dimension of the system since 
$\alpha_{3D}=\frac{|U_{BF}| (U_{BB}n_0)^{\frac{1}{2}}}{(2J_B)^{\frac{3}{2}}}$,
 while $\alpha_{2D}=\frac{|U_{BF}|}{2J_B}$, i.e. $\alpha$ is independent of the 
Bose-Bose interaction $U_{BB}$ and the condensate density $n_0$ in $D=2$.
The condition $\alpha \ll 1$ provides a constraint to the maximum value of the Bose-Fermi interaction 
where our theory can still safely applied. Indeed we have 

\begin{equation} 
\left\{ \begin{array}{ccc}
&|U_{BF}^{max}|=2J_B \ \   &\mbox{in}\ \  D=2 \\ 
&|U^{max}_{BF}| = \frac{(2J_B)^{\frac{3}{2}}}{(U_{BB}n_0)^{\frac{1}{2}}} \ \ &\mbox{in}\ \  D=3
\end{array} 
\right.
\end{equation}

Strictly speaking, however, the condition $\alpha > 1$ does not imply that the bosonic condensate cannot
be described anymore within the Bogoliubov approach but only that the distortion of the condensate wavefunction
due to the impurities is sizable and a  full solution of the GP equation in the presence of impurities is required.

In this sense we would expect that for $|U| > U_{BF}^{max}$ our theory still \emph{qualitatively} applies,
being however not anymore quantitatively accurate, if the condensate fraction $\phi$ of the mixture is  
close enough to 1. Whenever $\phi$ is instead much smaller than 1, the Bogoliubov modes are not anymore the
appropriate quasi-particles to describe the bosonic system and an alternative treatment is needed.  

As discussed in Section \ref{theory}, the Bogoliubov spectrum in our approach does not depend
on the impurity distribution and therefore the properties of the Bogoliubov modes
can be estimated by applying the Bogoliubov approach for the pure system.
The condensate density $n_0$ is in general unknown and has to be calculated 
self-consistently within Bogoliubov theory for a given density $n_B$ and temperature $T$.
This requires adding one more equation to our approach, i.e. the number equation of Bogoliubov
theory in the condensed phase \cite{vanoosten}
\begin{eqnarray}
\label{numbereq}
&&\ \ n_B=n_0+  \\ 
&&\!\!\!\frac{1}{N_s}\sum_{{\bf k}\in FBZ}^\prime
\left(\frac{\epsilon^*_{\bf k}+U_{BB} n_0}{\hbar
\omega_{\bf k}}N_{\mathbf{k}}(T) +
\frac{\epsilon^*_{\bf k}+U_{BB} n_0-\hbar\omega_{\bf k}}{2\hbar
\omega_{\bf k}}\right) \nonumber
\end{eqnarray}
which we solve numerically. As already discussed above, strictly in the homogeneous 
two-dimensional case, the Hohenberg-Mermin-Wagner theorem predicts that thermal fluctuations destroy the
 condensate for arbitrarily low-temperatures and Eq. (\ref{numbereq}) cannot be used for finite $T$
in $D=2$.

\subsection*{$\zeta$ parameter}
For $\alpha\ll 1$ and $\phi \approx 1$ the condensate in the presence of static impurities can be safely 
described within the Bogoliubov approach presented in Section \ref{theory}. Moreover within these approximations
the Hamiltonian description given in Eq. (\ref{h_hol_lftransf}) applies also 
to the case of mobile impurities. However, in order to obtain a simple expression for the
renormalization factor $S$ in the single impurity case, we had to assume that the fermionic
hopping $J_F$ is much smaller than the polaron shift $E_p$ ($\zeta \ll 1$),
\begin{figure*}[t!]
    \begin{tabular}{cc}
      \resizebox{80mm}{!}{\includegraphics{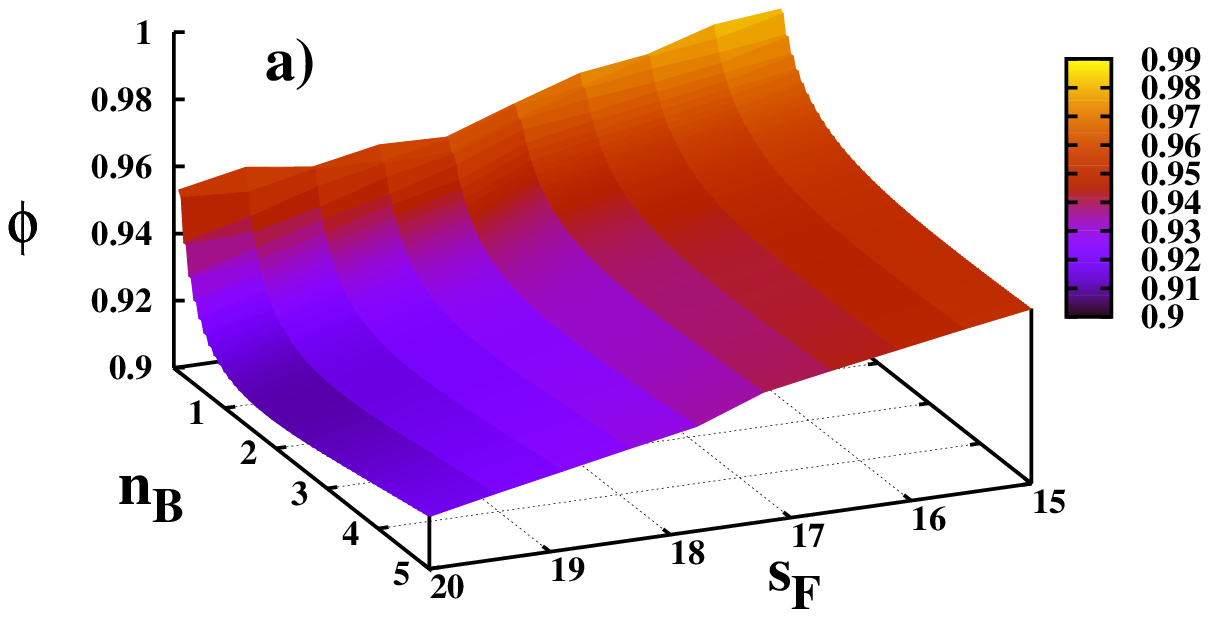}} &
      \resizebox{80mm}{!}{\includegraphics{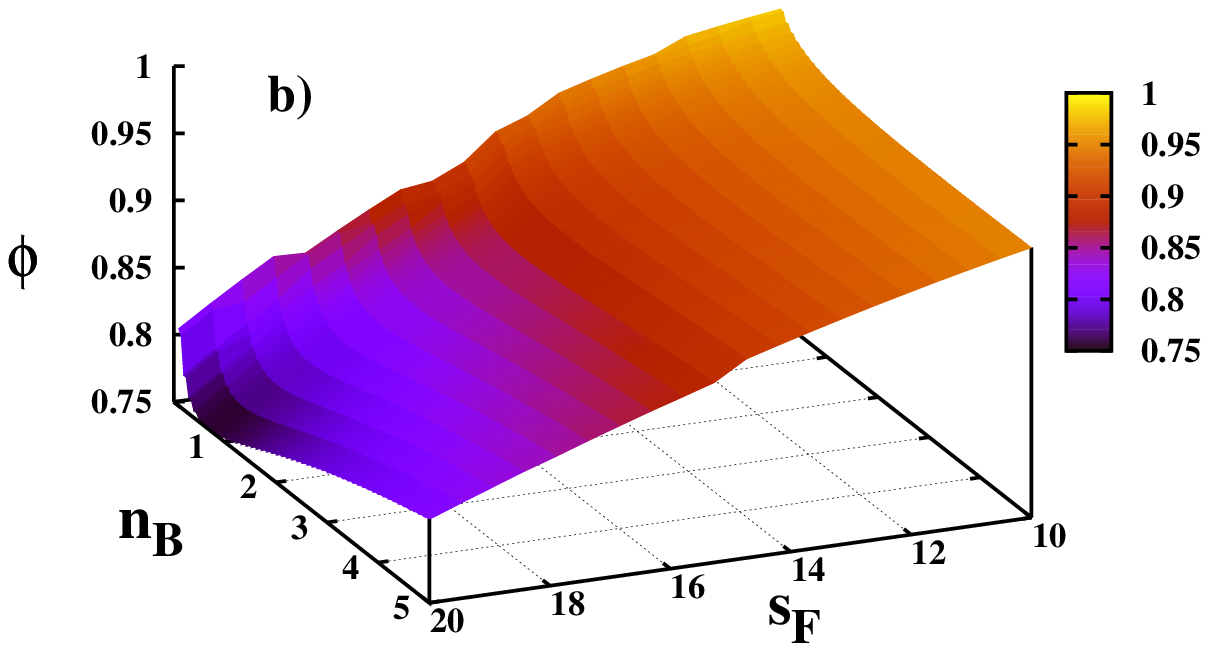}} \\
      \resizebox{80mm}{!}{\includegraphics{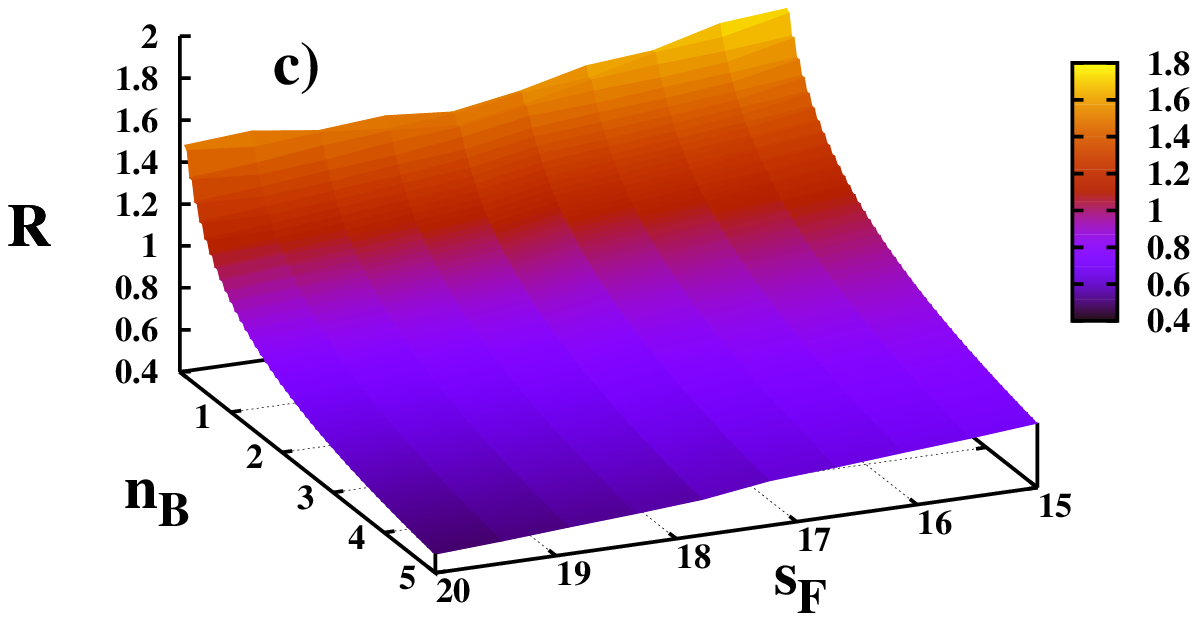}} &
      \resizebox{80mm}{!}{\includegraphics{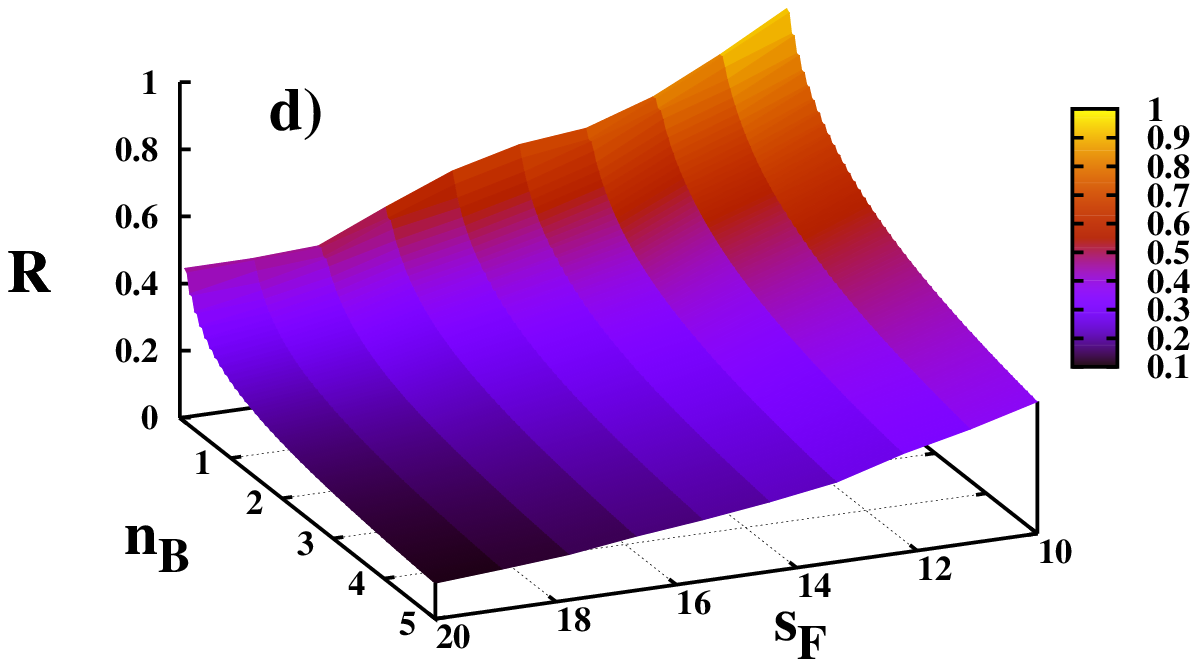}} \\
      \resizebox{80mm}{!}{\includegraphics{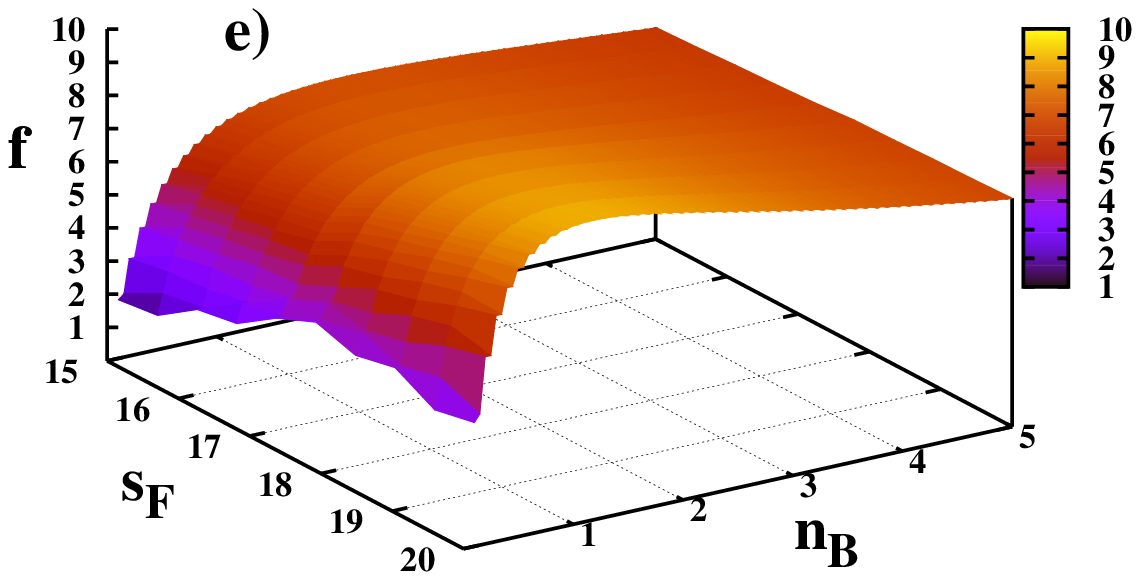}} &
      \resizebox{80mm}{!}{\includegraphics{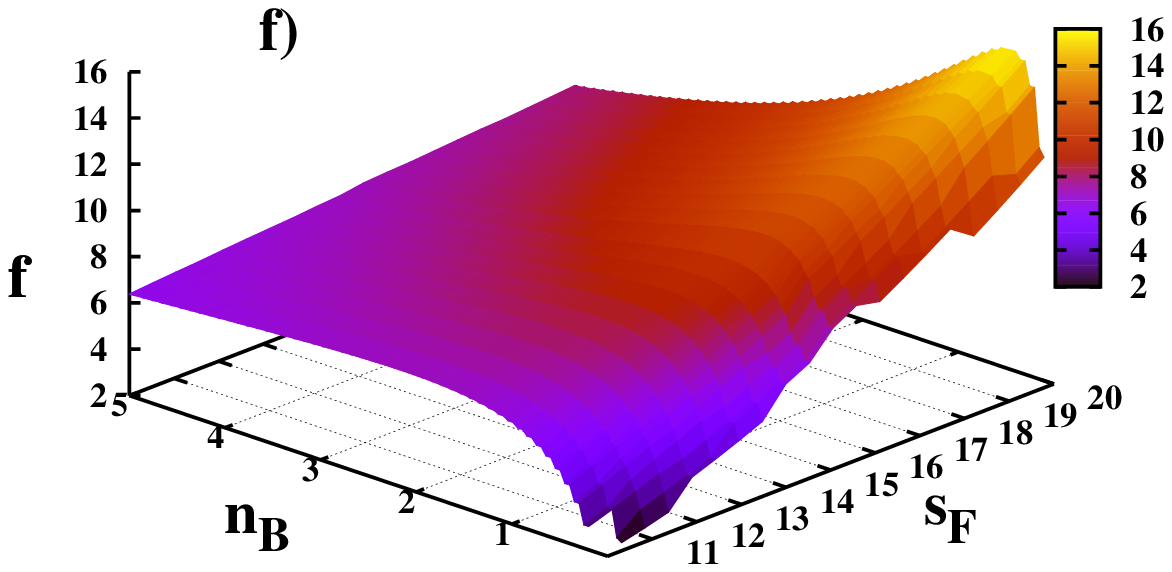}}\\
    \end{tabular}
    \caption{(Color online) (a-b) Condensate fraction $\phi$, (c,d) $R$ parameter,
 and (e,f) $f$ factor, defined in Eqs. (\ref{cond_frac},\ref{r_parameter}) and (\ref{f_factor}) respectively,
 for two three-dimensional $^{87}\rm{Rb}-^{40}\rm{K}$ setups with (a,c,e) $\gamma=1/3$ and (b,d,f) $\gamma=1/2$ 
as a function of the bosonic density $n_B$ and of the lattice depth $s_F$ at $T=0$. 
$f$ is expressed in units of $(E_r^F)^{-2}$. The minimum value of the lattice
depth $s_F$ is chosen such that $s_B\geq5$ in both cases.}
\label{3d_T0_summary}
\end{figure*}
 where $E_p$ has to be calculated from the theory.
According to the definition given in Eq. (\ref{polaron_shift}), the polaron shift is given by $E_p=U_{BF}^2 g$,
 where \begin{equation}
g=\frac{1}{N_s}\sum_{\mathbf{k} \in FBZ}
 \frac{n_0 \epsilon^*_{\mathbf{k}}}{(\hbar \omega_{\mathbf{k}})^2}
\end{equation}
 and only depends on the properties of the bosonic component. Therefore since the polaron shift increases with
the modulus of the Bose-Fermi interaction $U_{BF}$, the condition $\zeta \ll 1$ limits the minimum value of the Bose-Fermi
 interaction for which the formalism can be applied to $|U_{BF}| \gg |U_{BF}^{min}|=\sqrt{\frac{J_F}{g}}$

Motivated by the large number of parameters present in the theory, we summarize the range of parameters
 where our approach can be applied by introducing the ratio 
\begin{equation}
\label{r_parameter}
 R=|\frac{U_{BF}^{max}}{U_{BF}^{min}}|=|\frac{a_{BF}^{max}}{a_{BF}^{min}}|
\end{equation}
 such that for $R > 1$ there is a window
 of parameters where our approximations can be simultaneously satisfied. 
Intuitively the condition $\zeta \ll 1$ requires that the fermionic impurities
move much slower than the typical time taken by the phononic cloud to rearrange itself. Therefore 
in general small values of the fermionic hopping $J_F$ and $\gamma$ are more favorable to our approach,
meaning that mixtures where the bosons move faster than the fermions would be in general a better choice to 
reach the regime under investigation, even though the energy scales involved also crucially depend on the
interaction $U_{BB}$ and on the density $n_B$ of the condensate.

\section{Hopping renormalization in a three-dimensional $^{87}\rm{Rb}-^{40}\rm{K}$ mixture}
\label{results}
To be more concrete in this Section, we refer to the most commonly studied 
Bose-Fermi mixture, i.e. \hspace{-0.1cm}a $^{87}\rm{Rb}-^{40}\rm{K}$ mixture loaded into an optical lattice
in an experimental setup similar to the one used in Ref. [\onlinecite{best}].
For $^{87}\rm{Rb}-^{40}\rm{K}$ close to the interspecies Feshbach resonance, the bosonic scattering length
is $a_B \approx 100a_0 = 5.3 \rm{nm}$, where $a_0$ is the Bohr radius. Increasing values of 
the Bose-Bose scattering length are generally unfavorable to our approach, since this increases both $\alpha$ 
and $\zeta$ (Eqs. (\ref{alpha}) and (\ref{slow_imp})) and decreases the condensate fraction $\phi$ (Eq. (\ref{cond_frac})),
 if the other parameters stay unchanged. We found that smaller values
of the Bose-Bose scattering length substantially enlarge the range of parameters where our approach is 
quantitatively valid. This suggest that different mixtures with smaller $a_B$,
like the $^6Li-^{23}\rm{Na}$ mixture theoretically studied in [\onlinecite{tempere_prb2009}] 
($a_B \approx 53a_0 = 2,8  \rm{nm}$ for
$^{23}\rm{Na}$ \cite{samuelis}), could be even better candidates to realize the regime under investigation, if loaded into optical lattices.
Since, however, experimental data about lattice Bose-Fermi mixtures involving those species are not yet available to
the best of our knowledge, we decided in this work to focus on the $^{87}\rm{Rb}-^{40}\rm{K}$ mixture and postpone the analysis
of other mixtures to future publications.  

As explained in the previous section we only present results for the case $D=3$, and
we first focus on the $T=0$ case. Due to the large value of the Bose-Bose scattering length, 
we have to restrict our analysis to rather small values of the parameter $\gamma$, 
i.e. $\gamma=1/3,1/2$, where the bosons are substantially faster than 
the fermions \cite{will}. Results are summarized in Fig. \ref{3d_T0_summary}. 

As evident in Figs.~\ref{3d_T0_summary}a and \ref{3d_T0_summary}b for both setups 
the condensate fraction $\phi$ is relatively large and for $\gamma=1/3$ is always above $90\%$.
Small values of the lattice depth are in general favorable to the consistence
of Bogoliubov approach since for a given $a_B$ the bosonic system is less correlated.
However $s_F$ cannot be reduced at will in order to stay in a parameter regime where the Hamiltonian 
in Eq. (\ref{bosefermi_h}) applies. For $\gamma=1/2$ the quantum depletion of the condensate is much larger
and only for the lowest lattice depth under investigation ($s_F=10-12$) the condensate fraction is above 0.9.
Considering the parameter $R$ defined in Eq. (\ref{r_parameter}) in Figs.~\ref{3d_T0_summary}c 
and \ref{3d_T0_summary}d, it is evident that only for $\gamma=1/3$ there is a sizable window of parameters 
where our approach is quantitatively valid, i.e. $R >1$.  We however realized that the strongest limitations arise
from the condition $\alpha \ll 1$, which constrains the maximum value of the Bose-Fermi interaction,
rather than $\zeta \ll 1$. As discussed in the previous section, whenever  $\alpha \ll 1$ is violated but the 
Bogoliubov approach is still expected to apply, we expect our results to be qualitatively valid 
and therefore show the results for $\gamma=1/2$ in the plot for comparison. Low bosonic densities and shallow lattices 
are favorable to the theory. Indeed, by reducing the lattice depth $s_F$, the range of densities where the
theory applies is substantially increased.  
\begin{center}
\begin{figure}[htb]
   \includegraphics[scale=0.4,angle=-90]{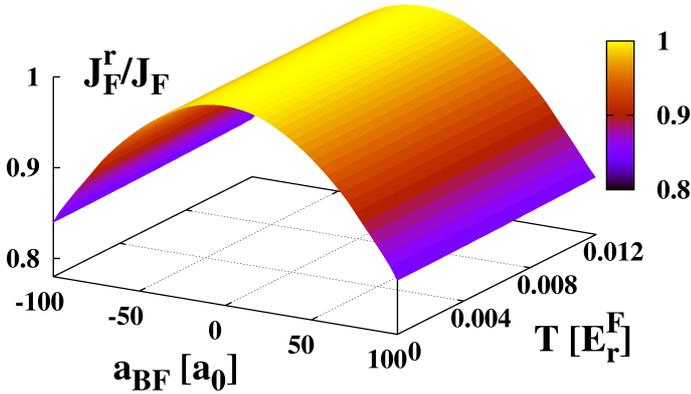}
    \caption{(Color online) Hopping renormalization $J_F^r/J_F$ for $n_B=0.1,\gamma=1/3$ and $s_F=15$
 as a function of the Bose-Fermi scattering length expressed in units of Bohr Radii $a_0$ and of
 the temperature $T$ expressed in units of fermionic recoil energy $E_r^F$.}
\label{finiteT}
\end{figure}
\end{center}
In Figs.~\ref{3d_T0_summary}e and \ref{3d_T0_summary}f we plotted results for the parameter 
 $f(n_0(n_B),s_F)_{|T=0}$ defined in Eq. (\ref{f_factor}), recalling that the renormalized hopping 
$J^r_F$ is related to $f$ by the simple relation $J^r_F=J_F e^{-U^2_{BF}f}$.
Both setups show a similar behavior of the $f$ parameter, though the numerical values are quite different for different setups
and the data for $\gamma=1/2$ are only shown for comparison since $R < 1$ in that case. For fixed (and small) values of the lattice
depth $s_F$, $f$ increases quite rapidly with the density $n_B$ and then saturates. For larger values of $s_F$ instead,
$f$ shows a maximum at intermediate densities and then slightly decreases. However the decrease at large $n_B$ and 
$s_F$ and the non-monotonic behavior could result also from a loss of accuracy in our approach, 
since in that region of parameters $R<1$ in both setups.

For finite temperature, we concentrate on a specific set of parameters, i.e. we choose $\gamma=1/3,s_F=15$ ($s_B=5$)
and low-density $n_B=0.1$, which is the regime where our approach quantitatively applies in a larger parameter range.
In a real experiment where the hopping renormalization is measured by observing the cloud expansion, the initial
configuration of the gas would be inhomogeneous due to the confining potential. This would mean that,
once the trapping potential is removed, different parts of the fermionic cloud would expand in the lattice
with a different renormalized hopping $J_F^r$ due to local value of the bosonic density $n_B$. We found that the renormalization factor 
$S$ in general increases with increasing density $n_B$, which would imply that the edge of the cloud, where $n_B$ 
is smaller, will expand faster. The results shown in Fig. \ref{finiteT} for the hopping renormalization $J_F^r/J_F$
 as a function of the temperature $T$ and of the Bose-Fermi
 scattering length $a_ {BF}$ are representative of the experimental situation at the edge of an expanding
 cloud and provide an upper bound for the renormalized hopping. The temperature range is chosen such
 that $T < T^{max}=E_p(U_{BF}^{max})$, where $E_p$ is the polaron shift, and $\phi > 0.9$.
We expect our estimate to be quantitatively more accurate for $40a_0 >|a_{BF}|> 60a_0$.

As evident from Fig. \ref{finiteT}, for fixed temperature 
the hopping renormalization as a function of $a_{BF}$ takes in our approach a Gaussian shape 
and the renormalized hopping decreases with the modulus of the Bose-Fermi scattering length,
such that $J_F^r/J_F\approx 0.95$ for $|a_{BF}|\approx 50a_0$ at the border of the bosonic 
cloud ($n_B=0.1$). A much larger renormalization effect is expected for larger densities.  

The effect of the temperature is very small in the range of parameters investigated and 
it is hardy visible in Fig. \ref{finiteT}. We found that for fixed $a_{BF}$
the renormalization factor $S$ slightly \emph{decreases} ($J_F^r/J_F$ increases) with increasing temperature, due to the dominant effect 
of the thermal depletion of the condensate. This trend is opposite to the one naively expected (see Fig. \ref{damping}) and also in contrast with the 
condensed matter case, where $S$ increases with $T$ \cite{holstein2}.
It is maybe worth mentioning that for higher densities (not shown) we found a nonmonotonic behavior in $J_F^r/J_F(T)$,
due to the increasing relevance of the thermal population of Bogoliubov modes at larger $T$. Since, however, large densities are
generally unfavorable to our approach, this effect could result as well from a loss of accuracy.     

\section{Conclusions}
\label{conclusions}
In this work we described the emergence of polaronic effects in Bose-Fermi mixtures in optical lattices.
Our approach is closely related to Ref. [\onlinecite{jaksch_njp1}] and is based on using the Bogoliubov
approach to describe the bosonic component of the mixture, considering first static and then
slowly moving fermionic impurities. The main difference to the case addressed in Ref. [\onlinecite{jaksch_njp1}]
is that in our case both species are substantially affected by the same optical lattice, as
 in currently available experimental setups. We showed that the effect of the optical lattice on the bosonic
condensate does not radically change the main conclusions for the homogeneous case \cite{bruderer_new}.
However the range of experimental parameters where the approach applies is substantially modified, 
whenever fermions and bosons move through the same optical lattice, since their
hopping parameters $J_F$ and $J_B$ are related to each other. 
  
For static impurities weakly coupled to the condensate ($\alpha \ll 1$), we have shown that 
an approximate treatment of the GP equation for the condensate wavefunction is possible and 
the Bogoliubov spectrum does not depend on the distribution of the fermionic impurities. However we expect 
the Bogoliubov treatment of the condensate in the presence of impurities to be still valid for larger $\alpha$ values,
provided the condensate fraction $\phi$ is large. Consequently, our results are expected to be qualitatively valid 
even beyond the regime $\alpha \ll 1$. We postpone a more quantitative treatment of the the regime with larger $\alpha$
and $\phi=\mathcal{O}(1)$, which would require a fully self-consistent GP + Bogoliubov approach, to a future work.

We derived an analytical expression for the hopping renormalization of a single impurity in the regime 
$\zeta \ll 1$ (slow impurity). This effect can be measured in experiments involving 
strongly imbalanced ($N_F/N_B \ll 1$) Bose-Fermi mixtures in optical lattice, by observing the expansion of the
fermionic cloud in the lattice when the trapping potential is suddenly removed \cite{bloch_polaron}. Within our approach, the fermionic hopping
in this regime is \emph{exponentially} renormalized due to polaron formation, i.e. $J^r_F=J_Fe^{-S}$.
In the relevant parameter range, the renormalization factor $S$ is found to be proportional to the square of the Bose-Fermi interaction.
Therefore we expect for $J^r_F$ a Gaussian dependence on the Bose-Fermi interaction $U_{BF}$ (or equivalently $a_{BF}$) and no dependence
on the sign of $U_{BF}$ ($a_{BF}$). This would lead to very strong experimental signatures of polaron physics once the considered regime
is reached.

The temperature dependence of the renormalization factor $S$ results from a competition between the thermal depletion of the condensate,
which induces a temperature dependence on the Bogoliubov spectrum $\hbar \omega_{\mathbf{k}}$ through the condensate density $n_0(T)_{|n_B}$,
and the thermal population of phononic modes. This dependence of the phononic spectrum on the temperature is missing in the standard
condensed matter case, where the renormalization factor $S$ always increases with $T$.  

In order to provide a better connection with experiments, we discussed the relevant parameter regime 
for a three-dimensional $^{87}\rm{Rb}-^{40}\rm{K}$ mixture in an optical lattice. Due to the large value of the bosonic scattering
length $a_B$, we have considered setups where $J_F \ll J_B$ and low bosonic densities. 
We found a sizable renormalization effect already for $n_B=0.1$. For a fixed value of the Bose-Fermi 
scattering length $a_{BF}$, the temperature dependence of the renormalized hopping is found to be opposite
to the one naively expected. The renormalization factor $S$ slightly \emph{decreases} for increasing $T$,
being dominated by the thermal depletion of the condensate rather than by the increasing phononic population.
This represent a mayor difference with the standard condensed matter case.

\section*{Acknowledgements} 
We thank S. Will and I. Bloch for useful discussions and first pointing out the possibility of measuring 
polaronic effects in lattice Bose-Fermi mixtures. A.P. thanks S. Knoop for providing 
interesting details about optical lattices and the Physics Department of Lund University (Sweden) 
for its kind hospitality during the completion of this work. This work was supported by the 
Deutsche Forschungsgemeinschaft DFG via Sonderforschungsbereich SFB-TR/49 and Forschergruppe FOR 801.


\begin{thebibliography}{99}
\bibitem{will} The condition $\gamma=1/3,1/2$ can be realized by choosing $\lambda \approx 764,762 \rm{nm}$. These values of the lattice
wavelength are however rather close to the atomic resonances, where a substantial heating effect from photon scattering 
is expected. This could make harder the experimental observation on long timescales.
\end{thebibliography}
\end{document}